# Proximity Matters: Analyzing the Role of Geographical Proximity in Shaping AI Research Collaborations

Mohammadmahdi Toobaee[1,2], Andrea Schiffauerova[1], and Ashkan Ebadi[1,3,*]

[1] Concordia Institute for Information Systems Engineering, Concordia University, Montreal, QC H3G 2W1, Canada
[2] Department of Quantitative Studies, University Canada West, Vancouver, BC V6Z 0E5, Canada
[3] Digital Technologies, National Research Council Canada, Toronto, ON M5T 3J1, Canada
[*] Corresponding author. E-mail: ashkan.ebadi@nrc-cnrc.gc.ca

**Abstract** The role of geographical proximity in facilitating inter-regional or inter-organizational collaborations has been studied thoroughly in recent years. However, the effect of geographical proximity on forming scientific collaborations at the individual level still needs to be addressed. Using publication data in the field of artificial intelligence from 2001 to 2019, in this work, the effect of geographical proximity on the likelihood of forming future scientific collaborations among researchers is studied. In addition, the interaction between geographical and network proximities is examined to see whether network proximity can substitute geographical proximity in encouraging long-distance scientific collaborations. Employing conventional and machine learning techniques, our results suggest that geographical distance impedes scientific collaboration at the individual level despite the tremendous improvements in transportation and communication technologies during recent decades. Moreover, our findings show that the effect of network proximity on the likelihood of scientific collaboration increases with geographical distance, implying that network proximity can act as a substitute for geographical proximity.

## 1. Introduction

Collaboration is a key driver of scientific output and performance (Ebadi & Schiffauerova, 2015, 2016). The critical role of collaboration in facilitating the production of new knowledge across different fields of science (Adams, 2013; Jones et al., 2008; Wuchty et al., 2007) and the positive effect of knowledge production on the long-term economic growth (Aghion & Howitt, 1992; Jones, 1995) have been studied thoroughly in recent years. Moreover, long-distance scientific collaborations, including international collaboration, have shown to be even more important as they provide higher-quality research productions (Adams, 2013; Adams et al., 2005; Narin et al., 1991). It might be expected that revolutionary developments in transportation and communication technologies could play a crucial role in facilitating collaborations between scientists who are geographically far away from each other, diminishing the effect of geographical distance on collaboration (Castells, 1996; Johnson & Mareva, 2002). Nevertheless, several recent studies show that, despite all technological developments, geography is still among the main determinants of collaboration (e.g., Bergé, 2017; Bignami et al., 2020; Morescalchi et al., 2015). In addition, despite the acknowledged importance of geographical distance in collaboration, its by-products, such as differences in national systems, make collaboration even more challenging (Lundvall, 1992). Considering the importance of long-distance collaboration in producing high-impact research and the negative effect of geographical distance on forming scientific collaboration, one may ask if other forms of proximity, instead of geographical proximity, can encourage scientific collaboration.

Collaboration requires creating a connection between researchers; hence, it can be considered a social process (Freeman et al., 2014; Katz & Martin, 1997). Connections may gradually form a social network, and social networks are the drivers of their own evolution over time (Jackson & Rogers, 2007). Therefore, potential network effects that influence the collaboration process should not be neglected. Although some studies analyzed the evolution of scientific collaboration networks (Almendral et al., 2007; Balland, 2012; Barabâsi et al., 2002; Maggioni et al., 2007; Newman, 2001; Wagner & Leydesdorff, 2005), empirical studies on the impact of network proximity on collaboration, and the interaction between geographical proximity and network proximity are scarce. If there is a substitutability pattern between geographical and network proximities, network proximity would partially compensate for the negative effect of geographical distance. In this case, enhancing the network proximity of distant researchers through long-distance collaborations could help create new long-distance connections.

As a highly evolving and interdisciplinary field, artificial intelligence (AI) is anticipated to contribute ~16 trillion USD to the global economy by 2030 (PwC, 2017). The technology's potential for economic growth and technological innovation has encouraged enormous public and private investments in the field. In Canada, AI investments are expected to contribute to Canada's economy by $66 billion by 2030 (Government of Canada, 2018). Motivated by the attributes of AI technology and to mitigate the risk of heterogeneity, in this study, we focus on the AI scientific ecosystem and will investigate the effect of geographical distance on scientific collaboration at the individual level. In addition, we will assess if network proximity can substitute geographical proximity to form long-distance collaborations. We will consider four data scenarios to study the research objectives: 1) focusing on collaborations among Canadian AI researchers, 2) expanding the scope to scientific collaborations in North America, i.e., Canada and the United States, 3) expanding data to cover European AI researchers, and 4) including all AI researchers in the dataset. This would enable us to compare findings at different levels.

This work contributes to the literature in at least four ways. First, we study the geo-pattern of scientific collaborations among researchers at the individual level, and not geographical regions or organizations. This will shed light on some critical yet uncovered aspects of scientific collaborations among individuals that may not be discernable when studying collaboration among geographical regions or organizations. Second, we start with studying the geo-pattern of scientific collaboration among Canadian AI researchers and then extend the domain of study to the United States, Europe, and the entire world. This comprehensive step-by-step analysis will provide a better insight into understanding the role of different dimensions of proximity in scientific collaboration. Third, this study covers AI-related publications from 2001 to 2019. As AI has been among the fastest-growing fields since 2000 (PwC, 2017), many scientific papers on AI were published after 2000. Therefore, this work covers a wide range of recent AI publications. Finally, we perform a deep analysis using both conventional and machine learning techniques to better assess the relationship between scientific collaboration and its determinants.

The remainder of the paper is organized as follows. Section 2 covers relevant literature about the notion of proximity and its role in explaining scientific collaboration. Data and methodology are discussed in Section 3. The results of the data analysis are presented and discussed in Section 4. Section 5 discusses the main findings of this research work and concludes. Limitations of the research and some future research directions are presented in Section 6.

## 2. Related Work

The concept of proximity is a useful framework for analyzing the determinants of collaboration (Cunningham & Werker, 2012; Kirat & Lung, 1999; Torre & Rallet, 2005). Bergé (2017) argued that different dimensions of proximity favour collaboration in two general ways: 1) proximity enhances the chance for potential partners to meet, and 2) it reduces the costs involved in the collaboration. Therefore, it increases the expected net benefits of the collaboration and augments the likelihood of its success. However, various dimensions of proximity may have different effects on collaboration. Balland (2012) studied the impact of different types of proximity on collaboration and found that geographical, organizational, and institutional proximities favour collaborations, while cognitive and social proximities do not play a significant role.

Among different dimensions of proximity, geographical proximity is the most common dimension studied in the literature (Knoben & Oerlemans, 2006). It is argued that face-to-face interaction is essential to transfer tacit knowledge and conduct research. Gertler (1995), for example, addressed the importance of distance between collaborating parties for the successful development and adoption of new technologies via interviews with advanced manufacturing technologies and concluded that co-location is crucial to facilitate face-to-face interactions, consequently enhancing the chance of productive collaborations. Rallet and Torre (1999) also confirmed the importance of physical proximity in forming innovative collaborations and argued that the transfer of tacit knowledge implies frequent face-to-face relations. Howells (2002) also discussed the relationship between knowledge and geography in the innovation process and highlighted the importance of geographical proximity in transferring tacit knowledge. In addition, face-to-face interaction can facilitate coordination, communication, and direct feedback (Beaver, 2001; Freeman et al., 2014), enhancing successful collaboration. Therefore, the geographical distance may lower the likelihood of successful collaboration by diminishing knowledge exchange opportunities through face-to-face contact and incurring more significant travel costs (Katz, 1994; Katz & Martin, 1997). In a more recent study, Catalini (2018) investigated the relationship between co-location and the rate, direction, and quality of scientific collaboration and found that co-location enhances the chance of co-publication by 3.5 times. Geographical proximity can also increase the probability of potential partners meeting in the first place via attending social events such as seminars and conferences linked to geographical distance (Bergé, 2017). For instance, Breschi and Lissoni (2009) showed that the social embeddedness of researchers decays with geographical distance since they have more knowledge about geographically closer partners. Therefore, we should expect the influence of geographical distance on collaboration to be negative. This fact has been evidenced in several studies focusing on different types of data such as co-authorship (Bergé (2017), Hoekman et al. (2009), Ponds et al. (2007)), co-patenting (Maggioni et al. (2007), Morescalchi et al. (2015)), and cooperation among research institutions (Scherngell & Barber (2009)).

Although economic geographers have emphasized the economic advantages of geographical proximity, they have pointed out that other dimensions of proximity besides geographical proximity are vital in understanding collaboration. They believe that the effect of geographical proximity on collaboration may not be evaluated in isolation as other forms of proximity could also be significant in explaining collaboration. Boschma (2005), for example, argued that geographical proximity per se is neither a necessary nor a sufficient condition for collaboration as it can facilitate interactive learning by strengthening the other dimensions of proximity. Frenken et al. (2009) provided an analytic framework for spatial scientometric research. They proposed to use the concept of proximity, which distinguishes physical proximity from other forms of proximity (e.g., cognitive, social, organizational, and institutional) as determinants of scientific interaction.

Due to the intuitive connection between social bonds and collaboration (Katz & Martin, 1997), the structure of scientific collaboration networks has been investigated in several studies (Almendral et al., 2007; Barabâsi et al., 2002; Fafchamps et al., 2010; Newman, 2001; Wagner & Leydesdorff, 2005). In addition, some network mechanisms which can encourage collaboration have been discussed in the literature. Carayol et al. (2019) examined the notion of triadic closure as the tendency of two indirectly linked nodes to connect. Compared with dyads, partners' negative behaviours are less expected in triads, as the third agent who serves the relation can punish it (Bergé, 2017). This structure can become a triadic closure, which is especially useful for international collaborations in which assessing the reliability of partners might be difficult. In other words, collaborating with a partner of a partner can be favourable since it reduces the risks of collaboration by limiting opportunistic behaviours (Bergé, 2017).

Homophily can also affect the forming of new collaborations. This feature has been studied in different contexts. Sociologists have shown that similarity is a force that drives the development of new links (Mcpherson et al., 2001). Blau (1974) investigated the relationship among physics scientists and concluded that having similar research interests and personal characteristics can positively affect research relationships. In a more recent study, Bergé (2017) argued that the collaboration network could trigger new connections through homophily. He indicated that partners in every successful scientific collaboration should have shared some similarities, e.g., the same research topic and, the same approach to research questions. Therefore, if two agents have links with a third agent, there is an excellent chance that they share some similarities, which can lead to future collaboration.

Researchers can rely on the network to find their potential partners as it is an authentic source of information (Gulati & Gargiulo, 1999). Considering the importance of time for researchers (Katz & Martin, 1997) and the increasing demand for scientific collaboration (Jones, 2009), this network function seems crucial for finding the right partners. Fafchamps et al. (2010) showed that when researchers are closer to each other in the collaboration network, they have more chances to access information about each other, enhancing the probability of future collaboration.

In terms of the effect of geographical and network proximities on scientific collaboration, different hypotheses can be proposed. First, the influence of geographical proximity and network proximity can be independent, meaning that the benefit of network proximity is homogeneous for all potential collaborators regardless of their geographical distance. This independence assumption can be the case only if geographical and network proximities influence collaboration via completely unrelated mechanisms. Since the mechanisms through which they affect collaboration (e.g., improving trust or facilitating searching for a future partner) are the same for geographical and network proximities, their interaction could not be independent (Bergé, 2017). Therefore, two contrary patterns could be discussed: 1) complementarity, and 2) substitutability. In the complementarity pattern, a higher level of network proximity will enhance the chance of collaboration when agents are geographically close. Bergé (2017) argued that complementarity can be the case, especially when agents have a "taste for similarity". However, in the substitutability pattern, network proximity is necessary for collaboration among two geographically apart agents since they are not subject to any other forms of proximity. But, if they are geographically close to each other, a high level of network proximity has less importance in encouraging collaboration (Bergé, 2017).

Several studies focused on the effect of geography and network proximity on collaboration. For example, Autant-Bernard et al. (2007) studied the impact of geographical and network proximities on the chance of future collaboration, and Maggioni et al. (2007) investigated the determinants of patenting activity. Both studies concluded that network and geographical proximities would positively affect future collaboration. Bergé (2017) examined the interaction between network and geographical proximities and showed that the effect of network proximity on future collaboration is moderated by geography. Moreover, he found a substitutability pattern between geographical and network proximities showing that network proximity mainly benefits distant collaborations.

Since the effect of geographical proximity on forming scientific collaborations at the individual level has not been addressed thoroughly in previous works, our main objectives in this study are 1) to investigate the effect of geographical distance on scientific collaboration at the individual level, and 2) to assess if network proximity can substitute geographical proximity to form long-distance collaborations at the individual level. Based on the defined objectives, two hypotheses are tested as follows: 1) To investigate the relationship between geographical proximity and scientific collaboration, we develop the following hypothesis: **Hypothesis 1.** *A higher level of geographical proximity enhances the likelihood of future scientific collaboration.* And, 2) To investigate the interaction influence of geographical proximity and network proximity on scientific collaboration, the following hypothesis is tested: **Hypothesis 2.** *Geographical proximity moderates the effect of network proximity on the likelihood of future scientific collaboration.*

## 3. Data and Methodology
### 3.1. Data
Data collection and preparation involved several steps. First, the publications' bibliographic data, including but not limited to title, abstract, keywords, date of publication, and author list, were retrieved from Elsevier's Scopus, filtering in research articles, conference papers, book chapters, and books published from 2000 to 2019. We only included publications for which both title and abstract were available. We used the ("artificial intelligence" OR "machine learning" OR "deep learning") search query to extract AI-related publications where at least one of the mentioned phrases appeared in the title/abstract of the publication or the keywords section. This resulted in 45,738 publications written by 153,720 authors from 162 different countries. We filtered the dataset based on the four data scenarios. For the first scenario, we included publications published by all Canadian authors (n=670 publications). In the second scenario, papers published by researchers solely from Canada and the United States were considered (n=7,180 publications). The third scenario expanded the second scenario to include European publications (n=20,508 publications). And finally, the last scenario covers entire publications in the main database (n=45,738 publications).

We followed Bergé (2017) approach and constructed dependent and independent variables in different time windows to prevent simultaneity biases. Thus, we considered 2-year and 3-year sliding time windows to calculate the dependent and independent variables, respectively. To form the dataset for each data scenario, we first created the co-authorship network that contained all possible scientific collaborations among authors who had at least one publication during the 2-year time window and at least one publication during the 3-year time window. A network was then created for each 2-year time window in which each node represented an author, and each edge represented a possible collaboration between two given authors and a dataset record. Then, dependent and independent variables were calculated. Co-publication, used as a proxy of scientific collaboration in several studies (Dahlander & McFarland, 2013; Hoekman et al., 2009; Ponds et al., 2007), was considered the dependent variable. This binary variable is equal to 1 if authors had at least one co-publication during the 2-year time window and 0 otherwise. Independent variables were calculated based on authors' publications during the respective 3-year time window. These steps were repeated for all 2-year time windows. We built the final dataset for each scenario with concatenating datasets resulting from each 2-year network. The following independent variables were calculated:

*1. Geographical proximity:* Following different studies (Bergé, 2017; Cunningham & Werker, 2012), we used direct spatial or "as the crow flies" distance between researchers' affiliations to measure their geographical proximity. For authors with more than one affiliation, we considered their first affiliation. To calculate the geographical distance between researchers, we first geocoded their location by turning their affiliation addresses into geographical coordinates using Google Geocoding API. Then, $distance_{i,j}$ as the measure of geographical proximity between author *i* and author *j* was calculated using the Haversine formula (Inman, 1835) as in equation (1).

$$distance_{i,j} = R \times 2 \times \text{atan2}(\sqrt{a}, \sqrt{1-a})$$
$$with\ a = \sin^2(\frac{\varphi_j - \varphi_i}{2}) + \cos\varphi_i \times \cos\varphi_j \times \sin^2(\frac{\lambda_j - \lambda_i}{2}) \qquad (1)$$

In the equation, $distance_{i,j}$ is the geographical distance between author *i* and author *j* in kilometres, $\varphi_i$ and $\lambda_i$ are, respectively, the latitude and the longitude of the first author's affiliation in radian, $\varphi_j$ and $\lambda_j$ are, respectively, the latitude and the longitude of the second author's affiliation in radian, atan2 is the 2-argument arctangent, and $R$ is the average radius of the earth (~ 6,373 kilometres).

*2. Institutional proximity:* Institutional proximity often refers to the collection of practices, laws, and rules defined by the geographical setting (Boschma, 2005). We defined this variable at two levels, i.e., province and country. In the first data scenario, we defined a binary variable as a proxy for institutional proximity, capturing the effect of the same province collaborations, which takes 0 for two given researchers if they are located in the same province, and 1 otherwise. In the second data scenario, we defined two binary variables as proxies for institutional proximity, one to capture the effect of the same province and the other to capture the impact of the same country. Similarly, they take 0 when two researchers are located in the same province/country and 1 otherwise. In the third and fourth data scenarios, we defined one binary variable to capture the effect of the same country following a similar definition.

*3. Network proximity:* We followed Bergé (2017) and used the "Total Expected Number of Bridging Paths (TENB)", as defined in equation (2), to estimate the network proximity of researchers. Bridging paths are seen as being a medium for network proximity. The main driver of the idea is that the more two researchers have bridging paths, the closer they will be with respect to the network, and, they will be more likely to engage in collaboration, thanks to network-based mechanisms.

$$TENB_{i,j} = \sum_{k} \frac{g_{i,k} \times g_{j,k}}{n_k} \quad (2)$$

In equation (2), $TENB_{i,j}$ is the total expected number of bridging paths between author $i$ and author $j$, $g_{i,k}$ is the total number of co-publications between author $i$ and author $k$, $g_{j,k}$ is the total number of co-publications between author $j$ and author $k$, $n_k$ is the total number of publications by author $k$. If $TENB_{a,b} > TENB_{a,c}$, it means that author $a$ and author $b$ are closer in the network with respect to their indirect connections than author $a$ and author $c$.

*4. Cognitive proximity:* The Latent Dirichlet Allocation (LDA) topic model (Blei et al., 2003) was used to estimate the cognitive proximity among authors. For this purpose, a vector of topics was first generated for each publication after performing several preprocessing steps, such as concatenating the title and abstract, removing punctuation marks, lowercasing words, lemmatizing, and removing stop words. Next, using a topic coherence metric (Röder et al., 2015), an optimal range for the number of topics was identified. It was found that nine is the optimal number of topics for our study. Therefore, a vector with nine elements was generated for each publication, with the value of each element showing the proportion of the respective topic in that specific publication. After estimating the topic vectors for all publications, the knowledge base vector of each author was calculated as the average of topic vectors of all their works published during the past 3-year window. For any given author to have the knowledge base vector, they must have at least one publication during that period. Finally, the $Cognitive\ distance_{i,j}$ as the measure of cognitive proximity between author $i$ and author $j$ was defined as in equation (3) where $s_i$ and $s_j$ are the knowledge base vectors of author $i$ and author $j$, respectively; and $corr(s_i, s_j)$ is the correlation between the knowledge base vectors of author $i$ and author $j$.

$$Cognitive\ distance_{i,j} = 1 - corr(s_i, s_j) \quad (3)$$

When two authors have the same knowledge base vectors, the correlation between their knowledge base vectors would be equal to 1; consequently, their cognitive distance is 0, implying the highest level of cognitive proximity. At the other extreme, when the correlation between the knowledge base vectors of two authors is equal to -1, their cognitive distance would be equal to 2, implying the lowest level of cognitive proximity.

*5. Regional contiguity:* Following Bergé (2017), we added two variables of regional contiguity to capture the effects of geography that are not seized by geographical proximity. These binary variables that capture the impact of collaboration between researchers from contiguous provinces/countries take 0 for two given researchers if they are in two contiguous provinces/countries and 1 otherwise.

### 3.2. Methodology
We built two classes of models: 1) conventional logistic regression, and 2) machine learning classifier. This allowed us to compare the outcomes, providing a better understanding of the relationship between the dependent variables/target and the independent variable/features.

*1. Conventional logistic regression:* The model is specified as in equation (4) where "Co-publication" is a binary variable that takes 1 if two researchers have at least one co-publication, otherwise 0; the $F$ function is the cumulative probability density function of the logit distribution; $\beta'$ is the vector of the coefficients, and $X$ is the vector of independent variables. Vector $X$ includes the natural logarithm of geographical distance, the natural logarithm of *TENB*, the binary variable representing institutional proximity, the cognitive distance, and the binary variable representing regional contiguity. The interaction between geographical distance and TENB was also added to vector $X$.

$$P(Co\text{-}publication = 1) = F(\beta'X) \qquad (4)$$

*2. Machine learning classifier:* The problem was formulated as a supervised machine learning classification problem with the binary variable of co-publication as the target variable. The independent variables explained in the previous section were used as features to predict future co-publication among researchers. The area under the curve (AUC) was used as the performance metric to evaluate the classification results. The dataset was split into training (90%) and test (10%) sets. We used the Synthetic Minority Over-sampling Technique (SMOTE) (Chawla et al., 2002) to address the imbalancedness issue. Finally, several machine learning classifiers, i.e., logistic regression, nearest neighbours, naive Bayes, support vector machines, random forest, and extreme gradient boosting, were built following a five-fold stratified cross-validation schema. A two-step hyperparameter tuning module was also implemented with a random search layer on top to find the approximate range for each parameter, followed by a more refined grid search to find the optimal value for the hyperparameters. To analyze the effect of each feature on predicting future scientific collaborations, the SHapley Additive exPlanations (SHAP) technique (Lundberg & Lee, 2017) was employed.

4. **Results**
**4.1. Descriptive Analysis**
Figure 1 compares the average geographical distance between all possible pairs of authors and the average geographical distance between pairs with at least one scientific collaboration. In the first data scenario, which covers collaborations among Canadian authors, there are 2,702 observations. The dataset is imbalanced as observations associated with pairs of authors with at least one co-publication form only 3% of total observations. Analyzing the statistics of geographical distance among authors is informative; although the maximum geographical distance in the dataset is more than 4,400 kilometres, the maximum distance for authors with a co-publication is only 817 kilometres. Moreover, 75% of all collaborations happened among authors whose affiliations were close geographically (less than 7 kilometres). Analyzing the statistics of variables in the other data scenarios shows the same pattern. In all scenarios, the average geographical distance among authors in the collaboration subset is shorter than the average geographical distance in the entire dataset. Moreover, the average TENB of the whole dataset in all scenarios is considerably lower than the average TENB for their respective collaboration subset, implying network proximity's critical role in forming scientific collaborations. Also, comparing the average and maximum values for cognitive distance among authors in the collaboration subsets with the same parameters for authors in the entire datasets clearly shows that cognitive proximity influences forming scientific collaboration. Besides, the average of different country binary variables shows that authors from the same country are more likely to collaborate. In the second data scenario, only 1% of total collaborations happened among authors from different countries. Although this figure is higher for the third (36%) and the fourth (40%) scenarios, it still shows that institutional proximity is essential in scientific collaboration.

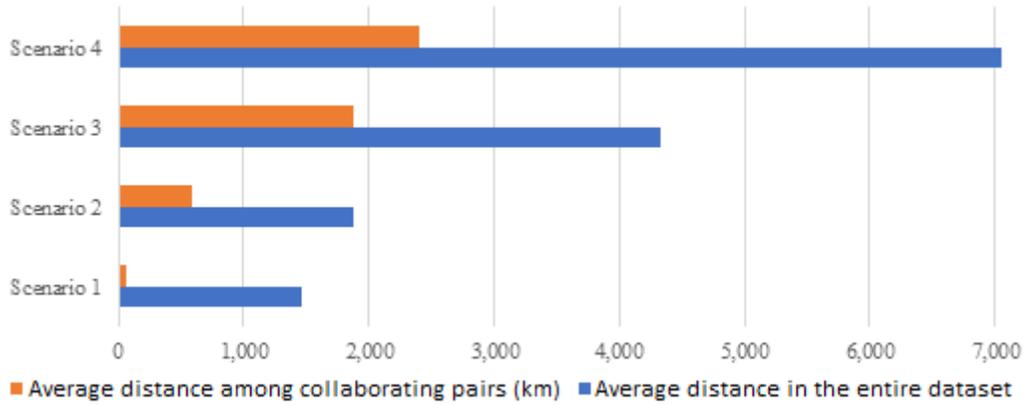

**Figure 1.** Comparing average distance among authors. Scenario 1) Canadian authors, Scenario 2) Canadian and U.S. authors, Scenario 3) Canadian, U.S., and European authors, and Scenario 4) all authors in the database.

### 4.2. Correlation Analysis

A negative correlation is observed between co-publication and geographical distance in all data scenarios (Figure 2). This may imply a negative relationship between geographical distance and scientific collaboration. In all scenarios, a strong and positive correlation between co-publication and TENB may indicate the critical role of network proximity in scientific collaboration. Besides, as expected, the correlation between co-publication and cognitive distance in all scenarios is negative. In the first two scenarios in which the binary variable of different provinces is included in the model, the negative correlation between co-publication and affiliation in different provinces may indicate the effect of institutional proximity on scientific collaboration. However, as expected, there is a positive and strong correlation between geographical distance, affiliations in different countries, and in different continents. For example, the correlation between geographical distance and affiliations in different continents in scenario 3 (Figure 2-c) and scenario four (Figure 2-d) are 92% and 80%, respectively. To avoid multicollinearity problems, we excluded the variables indicating different continent from those scenarios in further analyses.

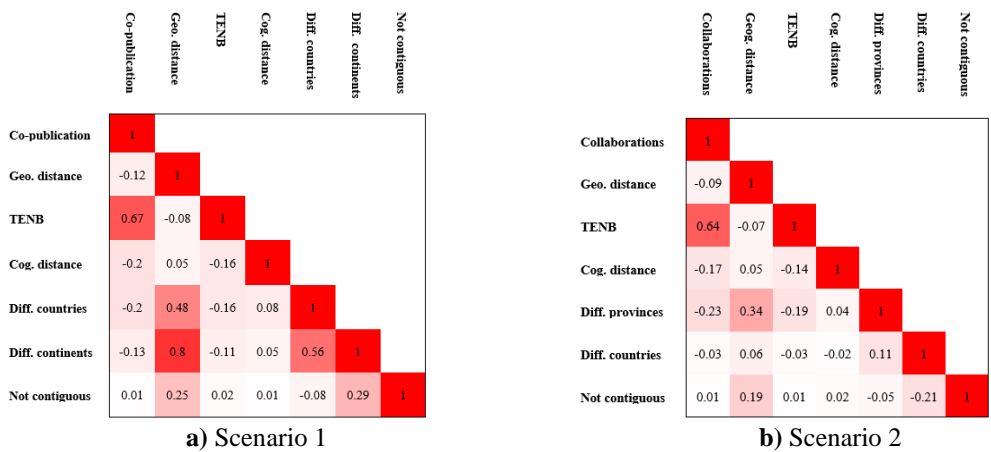

a) Scenario 1      b) Scenario 2

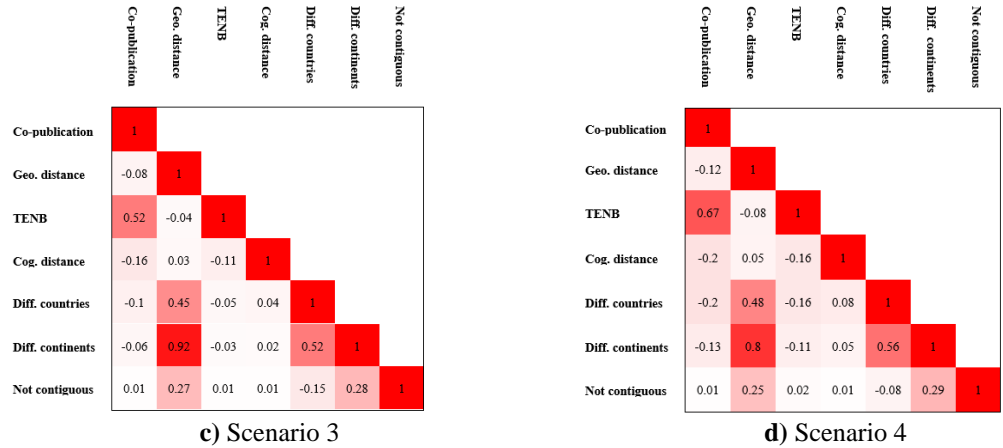

**Figure 2.** Correlation Analysis. **a)** Scenario 1, Canadian authors, **b)** Scenario 2, Canadian and U.S. authors, **c)** Scenario 3, Canadian, U.S., and European authors, and **d)** Scenario 4, all authors in the database.

### 4.3. Statistical Analysis

The logistic regression analysis results are reported in Table 1. In all scenarios, the negative and significant coefficient of geographical distance confirms the hindering effect of geographical distance on scientific collaboration. In addition, the positive effect of network proximity on scientific collaboration is reflected in the positive and significant coefficient of TENB. Besides, the coefficient of the interaction between geographical distance and TENB is positive and significant, implying that the effect of TENB on scientific collaboration is even more powerful when authors have a greater geographical distance. In all scenarios, the coefficient of cognitive distance is negative and significant, showing a lower likelihood of scientific collaboration for authors with farther research backgrounds. Also, the results show that ceteris paribus, being from different countries, would decrease the probability of collaboration among authors by 77%[1] to 29%[2].

**Table 1.** Logistic regression results. Scenario 1, Canadian authors, Scenario 2, Canadian and U.S. authors, Scenario 3, Canadian, U.S., and European authors, and Scenario 4, all authors in the database.

|  | **Scenario 1** | **Scenario 2** | **Scenario 3** | **Scenario 4** |
|---|---|---|---|---|
| **Dependent variable** | Co-publication | Co-publication | Co-publication | Co-publication |
| **Geo. distance (ln)** | -0.89*** (0.14) | -0.58*** (0.07) | -0.38*** (0.00) | -0.40*** (0.00) |
| **TENB (ln)** | 2.64*** (0.54) | 2.14*** (0.22) | 3.97*** (0.05) | 4.24*** (0.05) |
| **Geo. distance (ln) × TENB (ln)** | 0.44*** (0.15) | 0.66*** (0.04) | 0.41*** (0.01) | 0.43*** (0.01) |
| **Cog. distance** | -5.36*** (0.82) | -2.74*** (0.09) | -2.40*** (0.01) | -2.58*** (0.01) |
| **Different provinces** | 1.50 (0.87) | -0.49*** (0.43) | - | - |
| **Different countries** | - | -1.49*** (0.42) | -0.34*** (0.01) | -0.34*** (0.01) |
| **Not contiguous** | 1.00 (0.81) | 2.05*** (0.08) | 0.91*** (0.01) | 1.29*** (0.01) |
| **Number of observations** | 2,878 | 50,156 | 3,672,155 | 8,440,135 |
| **Pseudo R²** | 0.86 | 0.76 | 0.72 | 0.75 |
| **BIC** | 259.05 | 7,534.40 | 637,076.32 | 1,260,965.54 |

The standard errors are reported in parentheses. Coefficients marked with *** are significant at the 1% level.

---

[1] $(1 - \exp(-1.49))$

[2] $(1 - \exp(-0.34))$

## 4.4. Machine Learning Classifier

The machine learning classification results are presented in Table 2. Among all the classifiers we built, the Extreme Gradient Boosting (so-called XGBoost) provided the best AUC score in all data scenarios. In random search hyperparameter tuning, we considered 200 fits to find the approximate range of each parameter. Then, we fine-tuned the hyperparameters in a grid search with 7,290 fits. To assess the model's actual performance, the model was tested using the unseen test. The evaluation results of the best-performing mode on the unseen test dataset are presented in Table 3.

**Table 2.** AUC score on 5-fold cross-validation. Scenario 1, Canadian authors, Scenario 2, Canadian and U.S. authors, Scenario 3, Canadian, U.S., and European authors, and Scenario 4, all authors in the database.

|  | Scenario 1 | Scenario 2 | Scenario 3 | Scenario 4 |
|---|---|---|---|---|
| **Logit regression** | 0.98 | 0.97 | 0.95 | 0.96 |
| **Gaussian Naive Bayes** | 0.98 | 0.96 | 0.95 | 0.96 |
| **Nearest Neigbors** | 0.95 | 0.94 | 0.93 | 0.95 |
| **Support Vector Machine (SVM)** | 0.97 | 0.86 | 0.91 | 0.94 |
| **Random Forest** | 0.95 | 0.97 | 0.96 | 0.96 |
| **Extreme Gradient Boosting (XGBoost)** | **0.99** | **0.98** | **0.97** | **0.97** |

**Table 3.** AUC score on the unseen test dataset. Scenario 1, Canadian authors, Scenario 2, Canadian and U.S. authors, Scenario 3, Canadian, U.S., and European authors, and Scenario 4, all authors in the database.

|  | Scenario 1 | Scenario 2 | Scenario 3 | Scenario 4 |
|---|---|---|---|---|
| **Extreme Gradient Boosting (XGBoost)** | 0.94 | 0.95 | 0.90 | 0.91 |

Figure 3 demonstrates the beeswarm plot of SHAP values for different scenarios. This plot is designed to display an information-dense summary of how the features in a dataset impact the model's output. The features are sorted from the most important (on the top) to the least important (at the bottom).

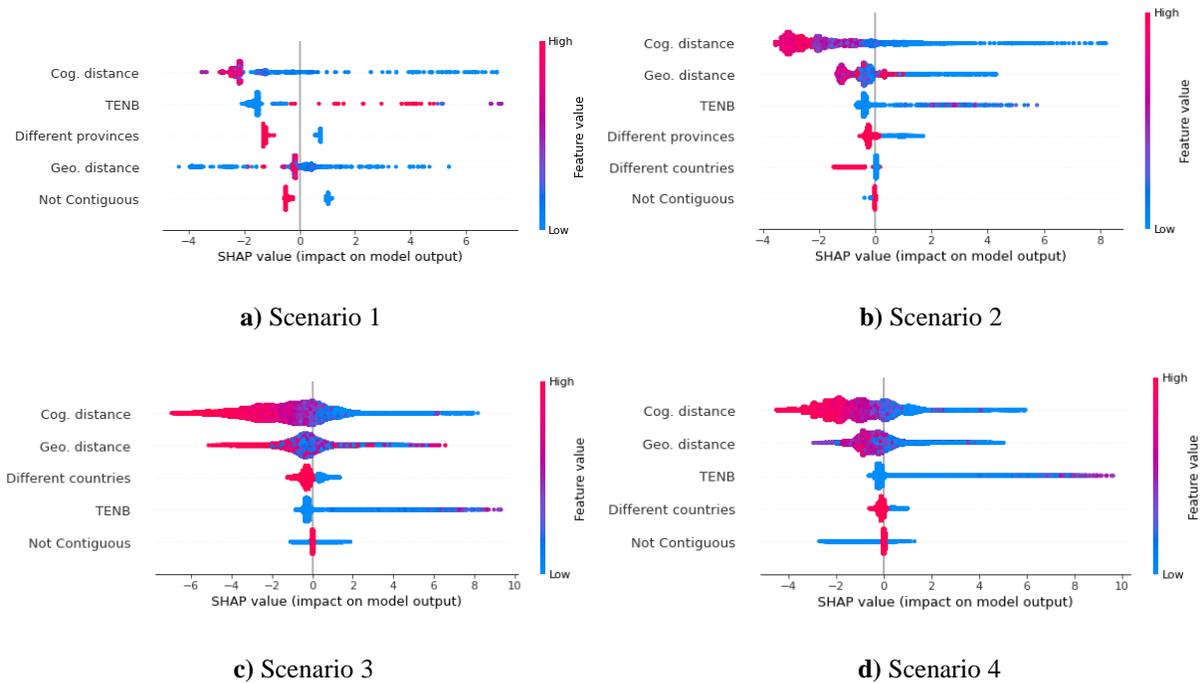

a) Scenario 1  b) Scenario 2

c) Scenario 3  d) Scenario 4

**Figure 3.** Beeswarm plots. **a)** Scenario 1, Canadian authors, **b)** Scenario 2, Canadian and U.S. authors, **c)** Scenario 3, Canadian, U.S., and European authors, and **d)** Scenario 4, all authors in the database.

As seen in Figure 3, cognitive distance has been the most important feature in predicting scientific collaborations in all studied scenarios. A high density of observations with a high level of cognitive distance on the left side of the diagram, which is associated with negative SHAP values, confirms the negative effect of cognitive distance on the likelihood of scientific collaboration. Regarding the geographical distance, although observations with close geographical distance are associated with both high and low SHAP values, the high density of observations with a high geographical distance on the left side of the diagram is discernible. The negative association between geographical distance and the chance of collaboration is more apparent in scenario 2 (Figure 3-b), scenario 3 (Figure 3-c), and scenario 4 (Figure 3-d).

## 5. Discussion and Conclusion

In the dynamic landscape of scientific research, the advent of AI has guided a new era of innovation and collaboration. Proximity could play a role in shaping research collaborations within the AI scientific ecosystem. For instance, with the proliferation of AI research hubs around the globe, understanding how geographical proximity influences the formation of research partnerships is of paramount importance. While digital communication and virtual interactions have seemingly diminished the role of physical distance, the intricate nuances of face-to-face interactions, knowledge exchange, and serendipitous encounters cannot be overlooked. By analyzing publication records spanning the years 2001 to 2019 within the realm of artificial intelligence, this study investigates the influence of geographical closeness on the probability of establishing forthcoming scientific partnerships among researchers. Furthermore, the study explores the interplay between geographical and network proximities to ascertain whether network proximity can act as a viable replacement for physical proximity in stimulating remote scientific collaborations.

We examined the relationship between the geographical proximity of researchers and the likelihood of their scientific collaboration. Using the co-publication data in the field of AI, we studied the geographical patterns of scientific collaboration among researchers in four scenarios. In the first scenario, we only included Canadian AI researchers. In the second scenario, AI researchers from the United States were added to the study. To have a more comprehensive understanding of geographical patterns of scientific collaboration, the scope of the study was extended to European countries and all countries around the world in the third and fourth scenarios, respectively.

Our statistical analysis results show a negative and significant association between geographical distance and the probability of scientific collaboration in all scenarios. All else being equal, a 10% increase in the geographical distance would decrease the chance of collaboration between authors by 3-7%. This result is also confirmed by the machine learning classification and SHAP feature importance analysis. As illustrated in Figure 3, observations related to authors with farther geographical distances are mainly associated with lower SHAP values, implying a lower chance of collaboration for those authors. These findings are in line with several previous studies (e.g., Bergé, 2017; Frenken et al., 2009; Maggioni et al., 2007; Morescalchi et al., 2015; Ponds et al., 2007; Scherngell & Barber, 2009) that found a negative association between geographical distance and scientific collaboration. Thus, we cannot reject the first hypothesis. Also, the regression results show a positive and significant relationship between network proximity and the chance of scientific collaboration in all scenarios. Ceteris paribus, a 1% increase in network proximity would enhance the opportunity for scientific collaboration by up to 7.2%. This finding is again confirmed by machine learning classification analysis. As seen in Figure 3, higher values of TENB are generally associated with higher SHAP values, implying that when authors are closer in the network, they are more likely to have at least one co-publication. This result aligns with Bergé (2017), which used the same measure, i.e., TENB, to gauge network proximity. In this study, two variables represented institutional proximity, i.e., different provinces included in the first and the second scenarios and different countries included in all scenarios except the first one. Although the coefficient of different provinces in scenarios 1 and 2 is not significantly different from zero, the machine learning feature importance analysis clearly shows an association between being from different provinces and a lower chance of scientific collaboration. However, the negative and significant association between different countries and the likelihood of scientific collaboration is much more discernible in the third and the fourth scenarios, where several countries are included. All else being equal, when authors are from different countries, their chance of collaboration would decrease by around 30%. This is confirmed by machine learning analysis, as observations related to authors from different countries are mainly associated with lower SHAP values. This finding confirms many previous studies (e.g., Bergé, 2017; Hoekman et al., 2009, 2010; Morescalchi et al., 2015; Scherngell & Barber, 2009) that found that being from different countries would negatively affect collaboration. The negative and significant coefficient of cognitive distance in all scenarios clearly shows that authors with higher cognitive distance are less likely to collaborate. However, the magnitude of this effect becomes smaller in scenarios that include more countries. The machine learning analysis also confirms the logistic regression results regarding the effect of cognitive distance on the likelihood of future scientific collaboration. According to Figure 3, cognitive distance is the most important feature in predicting future scientific collaborations in all scenarios. Besides, observations related to authors with higher cognitive distance are generally associated with lower SHAP values, implying a lower chance of scientific collaboration. This result is in line with several empirical studies that report the negative effect of cognitive distance on collaboration (e.g., Bergé, 2017; Cunningham & Werker, 2012; Ding, 2011; Jaffe & Hu, 2003; Jaffe & Trajtenberg, 1999).

The second hypothesis developed in this study was related to the substitutability of network proximity and geographical proximity. As the statistical analysis results show (Table 1), the coefficient of interaction between geographical distance and network proximity is positive and significant in all scenarios, implying that geographical proximity moderates the relationship between network proximity and the probability of scientific collaboration. In other words, as the effect of network proximity increases with geographical distance, network proximity does not seem to have a homogeneous overall impact. Instead, it acts as a substitute for geographical proximity. Figure 4 represents the estimated elasticity of the variable measuring the network proximity, TENB, on co-publications with respect to geographical distance, based on scenario 4 of Table 1. As seen, while network proximity has a positive impact on co-publication regardless of the geographical distance, its effect grows with distance, favouring the most long-distant scientific collaborations.

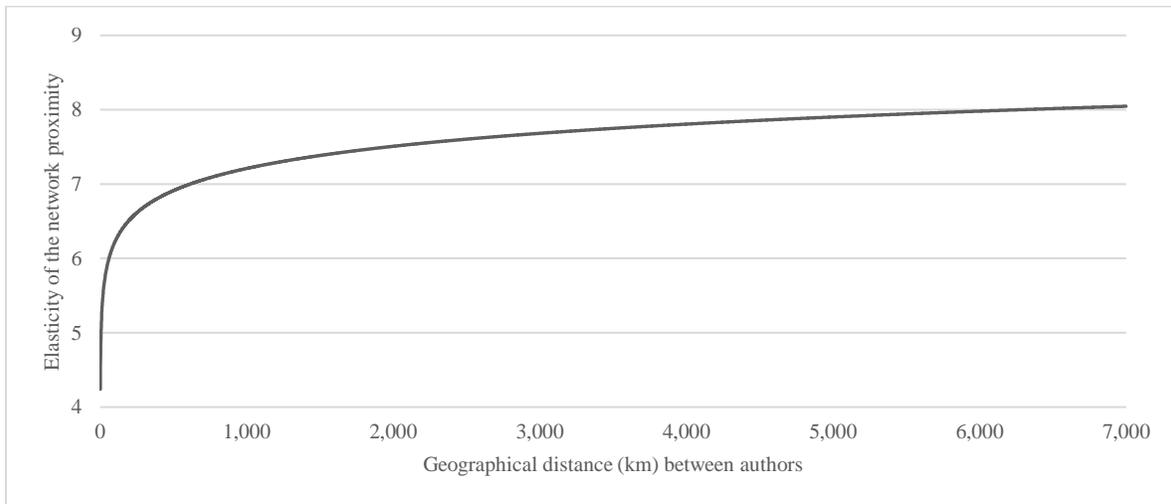

Figure 4. Interaction between network proximity and geographical distance

The substitutability pattern between geographical proximity and network proximity has an important implication from a policy-making point of view. Supporting long-distance scientific collaborations not only could result in higher quality research productions (Adams, 2013; Adams et al., 2005) but also may increase indirect connections among researchers, which in turn will trigger new scientific collaborations. Forming new long-distance collaboration increases the network proximity of researchers who had a scientific collaboration with the researchers in the new collaboration. This, in turn, may trigger new collaborations because of network effects, implying that more distant/more yielding collaborations are more likely to be established. In this sense, policies aiming at encouraging long-distance collaborations could positively affect knowledge production and ease future knowledge flows.

By delving into the interplay between geographical proximity and collaborative endeavors, in this study, we aimed to provide insights that can guide strategies for fostering impactful research collaborations, especially in the field of AI. The significance of this research lies in its potential to unravel the intricate dynamics that shape collaborative endeavors in the AI scientific landscape. As AI continues to drive innovation across various domains, understanding the role of geographical proximity in fostering research collaborations is crucial for optimizing collaborative strategies. By discerning whether network connections can compensate for geographical distances, the study addresses the evolving nature of collaboration in an increasingly interconnected world. The findings from this research can guide policymakers, funding agencies, and institutions in devising effective frameworks for nurturing collaborative environments within the AI field. Ultimately, by shedding light on the complex relationship between geographical and network influences, the study contributes to refining approaches that accelerate knowledge exchange, fuel scientific progress, and ultimately harness the full potential of AI-driven solutions.

## 6. Limitations and Future Work

This work has focused on scientific collaborations in the field of Artificial Intelligence. Thus, extending the study to other areas of science will be helpful to see whether they display the same pattern of substitutability between geographical proximity and network proximity. Besides, since we examined the determinants of scientific collaboration among individuals, adding individual-level control variables (e.g., age, gender, and number of publications) could provide a better understanding of the patterns of scientific collaboration. Furthermore, as the COVID-19 pandemic has led to enduring changes in the way humans collaborate across various domains, it becomes imperative to investigate the influence of different dimensions of proximity on future scientific collaborations in the post-pandemic era in upcoming research.